\documentclass[runningheads]{llncs}

\usepackage[T1]{fontenc}
\usepackage{graphicx}
\usepackage{multirow}
\usepackage{color}
\usepackage{tabularray}
\usepackage{adjustbox}
\usepackage[table,xcdraw]{xcolor}
\usepackage{rotating}
\usepackage{booktabs}
\usepackage{arydshln}
\usepackage{ragged2e}
\usepackage{tabularx}
\usepackage[hidelinks]{hyperref}
\usepackage{orcidlink}
\usepackage{bbding}
\usepackage[subtle]{savetrees}

\begin{document}
\title{SimBank: from Simulation to Solution in Prescriptive Process Monitoring}
\titlerunning{SimBank: from Simulation to Solution in PresPM}
\author{Jakob~De~Moor\textsuperscript{\Envelope}\inst{1}\orcidlink{0009-0006-4788-5346} \and
Hans Weytjens\inst{1,2}\orcidlink{0000-0003-4985-0367} \and
Johannes De Smedt\inst{1}\orcidlink{0000-0003-0389-0275}\and
Jochen De Weerdt\inst{1}\orcidlink{0000-0001-6151-0504}}
\authorrunning{J. De Moor et al.}
\institute{
    Research Centre for Information Systems Engineering, KU Leuven, Belgium \\
    \email{\{jakob.demoor,hans.weytjens,johannes.desmedt,jochen.deweerdt\}@kuleuven.be} \\
    \and
    School of Computation, Information and Technology, TUM, Germany \\
    \email{hans.weytjens@tum.de}
}

\maketitle
\begin{abstract}
Prescriptive Process Monitoring (PresPM) is an emerging area within Process Mining, focused on optimizing processes through real-time interventions for effective decision-making. PresPM holds significant promise for organizations seeking enhanced operational performance. However, the current literature faces two key limitations: a lack of extensive comparisons between techniques and insufficient evaluation approaches. To address these gaps, we introduce SimBank: a simulator designed for accurate benchmarking of PresPM methods. Modeled after a bank’s loan application process, SimBank enables extensive comparisons of both online and offline PresPM methods. It incorporates a variety of intervention optimization problems with differing levels of complexity and supports experiments on key causal machine learning challenges, such as assessing a method’s robustness to confounding in data. SimBank additionally offers a comprehensive evaluation capability: for each test case, it can generate the true outcome under each intervention action, which is not possible using recorded datasets. The simulator incorporates parallel activities and loops, drawing from common logs to generate cases that closely resemble real-life process instances. Our proof of concept demonstrates SimBank's benchmarking capabilities through experiments with various PresPM methods across different interventions, highlighting its value as a publicly available simulator for advancing research and practice in PresPM.

\end{abstract}

\section{Introduction}\label{sec:introduction}
Prescriptive Process Monitoring (PresPM) offers a valuable opportunity for businesses to enhance decision-making. Its methods leverage event log data from business processes and use machine learning to automatically prescribe interventions, such as machine maintenance, customer calls to maximize turnover, or loan application cancellations \cite{quovadis2021}. The goal is to optimize the specified target variable(s), such as delivery time (of an order), acceptance (of a loan offer), recovery (of a patient), availability (of a product), or a default rate (in manufacturing). The benefits of these prescriptive systems include product and service quality improvements, cost savings, and increased staff satisfaction. 

Despite their potential, significant limitations are present in current PresPM methodologies, particularly in benchmarking practices. Firstly, there is a notable lack of comprehensive comparisons among PresPM approaches. To the best of our knowledge, only one study has thoroughly compared multiple methods~\cite{weytjens2023}, though it only examines two. Additionally, existing research predominantly focuses on a limited range of intervention types, neglecting the development and testing of effective approaches for more complex intervention optimization problems, such as optimizing multiple sequential interventions. Secondly, accurate method evaluation remains a persistent challenge. Most evaluations rely on historical datasets~\cite{bozorgi2023RL,branchi2022,bozorgi2023CI,leoni2020,weinzierl2020}, which lack counterfactual outcomes, i.e. the results of actions not taken or recorded. Method performance has to be estimated, since not every prescribed action for a test case has been observed. Moreover, these estimates may be misleading as the (test) datasets reflect the policies under which they were collected (e.g., a bank’s loan approval criteria), yielding reliable performance estimates only for frequently observed parts of the optimization space.

To address these challenges in benchmarking and evaluation, we propose SimBank, a first-of-its-kind synthetic data generator designed specifically for PresPM research. SimBank simulates cases inspired by a bank’s loan application process and addresses the two main limitations in existing methodologies. First, SimBank ensures accurate method evaluation by providing a controlled environment where the underlying process-generating mechanisms are explicitly known. This allows for precise calculation of outcomes based on prescribed actions, eliminating the need to rely on historical data. Second, SimBank facilitates comprehensive comparisons between PresPM methods by enabling controlled experimentation. Researchers can adjust process parameters to simulate diverse scenarios, supporting both offline training (with pre-generated event logs) and online training (in dynamic, real-time environments).
Moreover, SimBank incorporates three main prescriptive intervention optimization problems, making it possible to compare methods across varying complexity levels. By combining these intervention types, SimBank enables testing of sequential intervention strategies---an area that remains underexplored in PresPM research. The simulator, based on commonly used logs, generates cases that closely reflect real-world process flows by including activities that occur in loops and in parallel. To validate SimBank's benchmarking capabilities, we conduct experiments with four different PresPM methods across the defined intervention scenarios. These experiments assess whether SimBank provides meaningful comparisons, enabling researchers and practitioners to accurately identify scenarios where specific PresPM methods excel or underperform. This approach bridges the gap between simulated and real-world environments, supporting the practical deployment of PresPM solutions.

The paper is structured as follows. Section \ref{sec:background} covers the background. Section \ref{sec:development} details the simulator development. Section \ref{sec:poc} presents a proof of concept to validate SimBank's effectiveness in comparing PresPM methods. We conclude this paper and suggest avenues for future work in Section \ref{sec:conclusion}.

\section{Background}\label{sec:background}
\subsection{Preliminaries}
PresPM is an emerging extension of Process Mining, which is dedicated to discovering, monitoring, and improving real-life process models~\cite{vanderAalst2009}. PresPM advances this goal by recommending case-specific interventions, aiming to directly optimize process outcomes. Recommendations are often derived from observational event logs, making them particularly useful in situations where traditional methods like randomized controlled trials (RCTs) or A/B testing are too costly or risky, such as a bank testing a loan assignment strategy. While still in its infancy, the field is steadily gaining momentum~\cite{branchi2022,bozorgi2023CI,quovadis2021,shoush2024}. 

Building on the framework in \cite{weytjens2023}, interventions in PresPM can be characterized along two dimensions: action width and action depth. Action width refers to the number of possible actions during an intervention, while action depth indicates the number of possible timing points for the intervention. When multiple interventions are performed in sequence, they form an intervention sequence. Table \ref{tab:dimensions} illustrates these dimensions with examples from marketing interventions aimed at increasing turnover through customer contact. These dimensions are instrumental in defining the complexity of an intervention.
\begin{table}
\caption{Intervention dimensions and examples}\label{tab:dimensions}
\setlength{\tabcolsep}{5pt}
\fontsize{7}{7}\selectfont
\centering
\begin{tabularx}{\textwidth}{X c c}
    \toprule
    \textbf{Example: contact a customer} & \textbf{Action width} & \textbf{Action depth}\\
    \midrule
    Intervene with a \textit{call} or \textit{email} at a given point in the process. & {2} & {1}\\
    \arrayrulecolor{gray}\midrule[0.25pt]
    Intervene with a \textit{call}, \textit{email} or \textit{visit} and decide at which of 4 given points in the process to intervene. & {3} & {4}\\
    \arrayrulecolor{gray}\midrule[0.25pt]
    Intervene with a \textit{call} or \textit{email} and decide at which of 2 given points in the process to intervene. Next, intervene with a \textit{visit} or \textit{post card} and decide at which of 4 given points in the process to intervene (intervention sequence). & {2 $\rightarrow$ 2} & {2 $\rightarrow$ 4}\\
    \arrayrulecolor{black}\bottomrule
\end{tabularx}
\end{table}

In this study, we focus on optimizing interventions for individual process instances in isolation, an assumption that simplifies the problem by disregarding interdependencies between cases. In real-world scenarios, such interdependencies---often due to shared resource constraints---can significantly impact outcomes. Recent research has begun to address these complexities \cite{padella2024,shoush2024}, but they remain outside the scope of this work---a simplification also made in the majority of existing studies. 

\subsection{Prescriptive Process Monitoring Approaches}
The first major approach adopted in PresPM is Causal Inference (CI), which focuses on estimating the effect of an intervention action on a target variable for each individual case using offline data, whether synthetic or real \cite{imbensrubinsCI}. A core challenge in CI is that only the outcome of the chosen intervention action is recorded for each case, while the counterfactual outcome remains unknown. One way to estimate this counterfactual outcome is to analyze similar cases in the data that received a different intervention action. However, real-world data-gathering policies often introduce confounding, where other variables influence both treatment/intervention and outcome. This can lead to treated and untreated cases differing strongly in their distributions, unlike the balanced and confounding-free setup of RCTs. Traditional CI methods attempt to correct for such distortions \cite{ivCI}. 
An example of CI in PresPM is the method described in \cite{bozorgi2023CI}, which incorporates CI by augmenting logs with estimated counterfactuals to identify optimal causal effect estimators. RealCause, a generative AI framework originally used as a CI evaluation tool, facilitates this process. 
Other studies include \cite{shoush2021}, \cite{shoush2022}, \cite{bozorgi2023RL}, and \cite{shoush2024}, although CI is not necessarily the main focus in the latter two.

Reinforcement Learning (RL) is the other most commonly adopted approach in PresPM. In RL, an agent learns to make (sequential) decisions by interacting with an environment, observing states, taking actions, and receiving rewards to develop a policy that maximizes cumulative rewards over time. A popular variant is Q-learning, where state-action values are collected in a Q-table, defining the policy \cite{sutton2018}. Deep Learning (DL) is frequently integrated with RL in applications with large state spaces, exemplified by Deep Q-learning (DQN), where a neural network approximates the Q-table \cite{dqn2013}. Online RL trains in real time, posing risks with its trial-and-error approach, but, like RCTs, avoids confounding. In contrast, offline RL uses potentially confounded pre-collected data, but is gaining popularity as a safer alternative.
Examples of online RL in PresPM include studies \cite{metzger2023} and \cite{shoush2024}, though real-life event logs are used to create the online environment, making their methods akin to offline RL despite online agent activity. Offline RL examples include \cite{branchi2022}, where similarity-based approaches are used to create training environments, and \cite{bozorgi2023RL}, where RealCause is applied to estimate counterfactual results.

\subsection{Limitations in Prescriptive Process Monitoring}\label{subsec:background_limitations}
Current approaches in PresPM mainly rely on historical offline datasets, with BPIC12 and BPIC17 \cite{bpic12,bpic17} being by far the most commonly used in works such as \cite{bozorgi2023CI,bozorgi2023RL,shoush2024,branchi2022}. Although insightful, historical datasets lack counterfactual outcomes. While we can estimate the impact of interventions, their true effects remain unknown, resulting in only approximate evaluations. RealCause is the most widely used evaluation method to obtain these approximate evaluations \cite{bozorgi2023CI,bozorgi2023RL,shoush2024,shoush2022,shoushwhitebox}. For example, in \cite{bozorgi2023RL}, RealCause not only generates alternative outcomes to train an RL agent, but also defines the evaluation metrics, assuming that these generated outcomes are the true outcomes.
Furthermore, comparison studies of PresPM methods are scarce. To the best of our knowledge, \cite{weytjens2023} is the only work to compare two implementations comprehensively by using a synthetic setup for precise evaluation. This scarcity is partly due to the lack of accurate evaluation mechanisms and unclear intervention dimensions, making comparisons difficult. These dimensions, i.e., action width and depth, are crucial to know, as the effectiveness of PresPM methods depends on them. For example, \cite{weytjens2023} shows that their CI implementation, though safer, performs worse than RL in scenarios with action depths larger than 1 (timing choice). It is also worth noting that most research focuses on interventions with action width 2 and action depth larger than 1 \cite{bozorgi2023CI,shoush2022,shoush2024}. For example, in BPIC12 and BPIC17, the act of sending multiple offers is considered an intervention and follows these dimensions. Optimization of intervention sequences is underexplored. Only the methods in \cite{weinzierl2020} and \cite{leoni2020} address this by deploying multiple predictive models. However, their approaches may accumulate errors, leaving room for improvement.
Lastly, commonly used logs often lack empirically validated causal relationships. Researchers typically select interventions without external domain expert confirmation, especially regarding their heterogeneous effects across different cases. For instance, while the intervention in BPIC12 and BPIC17 is correlated with higher loan acceptance rates, its causal impact has not been extensively studied or externally validated.

Manually defined simulators offer a solution to these limitations. In such a simulator, the process-generative function is fully known by design, allowing for the calculation of outcomes for all possible interventions, thus supporting accurate evaluation by generating counterfactuals. This setup also enables controlled experimentation, e.g., parameters such as confounding level can be adjusted systematically. Fully specifying the simulator from the ground up provides the flexibility to clearly define a range of interventions with varying complexity within the same process, and explicitly model their causal effects on outcomes. Combining these interventions also supports research for intervention sequences. 

\subsection{Current Simulations in Process Mining}
Given the benefits of a manually defined simulator for PresPM, it is important to understand why such a new simulator may be preferred over existing simulators. While CI and RL studies outside of Process Mining often use simulations~\cite{bica2020,sutton2018}, these do not generate business process data. Simulators have also gained traction within Process Mining. For example, Data-Driven Process Simulation (DDPS) approaches include \cite{camargo2020,kirchdorfer2024,pnsim2021}. Others train and employ DL models to generate simulated event logs, such as in \cite{camargo2019}. Examples of approaches that combine both, or hybrid simulations, are \cite{camargo2023,meneghello_simulator2024}.
Extending these simulations to PresPM research would be useful, as they can be applied to any event log and could enable realistic experiments. However, they have a major limitation in this context: these simulations are based on historical datasets, and thus still suffer from the same counterfactual limitations. They are typically developed to reflect the policy that generated the original event log (e.g., bank policy), making them unlikely to accurately generate outcomes when a PresPM method recommends actions that differ from past practice. Additionally, they do not support multiple intervention types or allow controlled experiments relevant to PresPM research, such as manipulating confounding levels. Even if certain components are manually defined to extend these simulations, the evaluation accuracy remains uncertain due to the limitations in the parts that are not manually defined but approximated.

In contrast, manually defined simulators provide the opposite trade-off: they offer accuracy in evaluation and enable fully customizable experiments, including support for multiple interventions and confounding scenarios. However, they are not automatically generalizable to any event log and may lack realism. Currently, only one such simulator exists for PresPM~\cite{weytjens2023}, but it is overly simplistic, lacking loops, parallelism, and a variety of interventions. This reveals a clear gap and opportunity for developing a more sophisticated, realistic specified simulator tailored to PresPM needs that provides customizable and fully accurate experiments.

\section{Simulator Development}\label{sec:development}
In this section, we introduce the development of SimBank. We begin by outlining the high-level requirements derived from our review of the literature, followed by a description of the process and its interventions. We then delve into the implementation details, highlighting SimBank’s capabilities.

\subsection{Requirements}
As discussed in Sections \ref{sec:introduction} and \ref{sec:background}, current research in PresPM faces two major limitations: a) challenges for accurate model evaluations, and b) an absence of comparisons of PresPM methods, both in evaluating them against each other and in assessing their effectiveness for different intervention types. To address these issues, we identified four requirements for SimBank:
\begin{itemize}
    \item \textbf{Complete evaluation.} The simulator should be able to calculate the target variable of a case for every possible action of an intervention. This overcomes the limitations of historical data in evaluating PresPM models, which lack counterfactuals. Moreover, it would allow for a direct comparison between true method performance and performance as approximated by commonly used evaluation techniques in PresPM and CI research, such as RealCause \cite{realcause2020}.
    \item \textbf{Inter-intervention method comparison.} The simulator should allow for a comparison of methods across various interventions with differing dimensions. There is significant potential for further research on intervention sequences, and our simulator should enable exploration of these more advanced optimization challenges.
    \item \textbf{Intra-intervention method comparison.} The simulator should support experiments within each intervention type to assess method performance across varying dataset characteristics (e.g., level of confounding). It should also accommodate methods requiring offline or online training, as both are commonly used (see Section \ref{sec:background}).
    \item \textbf{Balancing realism and simplicity.} A simulator for generating realistic datasets must balance complexity and interpretability. Business processes often include parallel activities, loops, and a high degree of structural variety, reflected in the diversity of trace variants \cite{schreiber2024}. Additionally, a simulator should create realistic variables relevant to its business context, e.g., a loan application process. At the same time, the generated datasets should remain simple enough to allow clear evaluation of PresPM methods. Too many dimensions or variables can obscure conclusions.
\end{itemize}

\subsection{Description of the Process and Interventions}
\begin{figure}
    \centering
    \includegraphics[width=1\textwidth, height=0.65\textwidth]{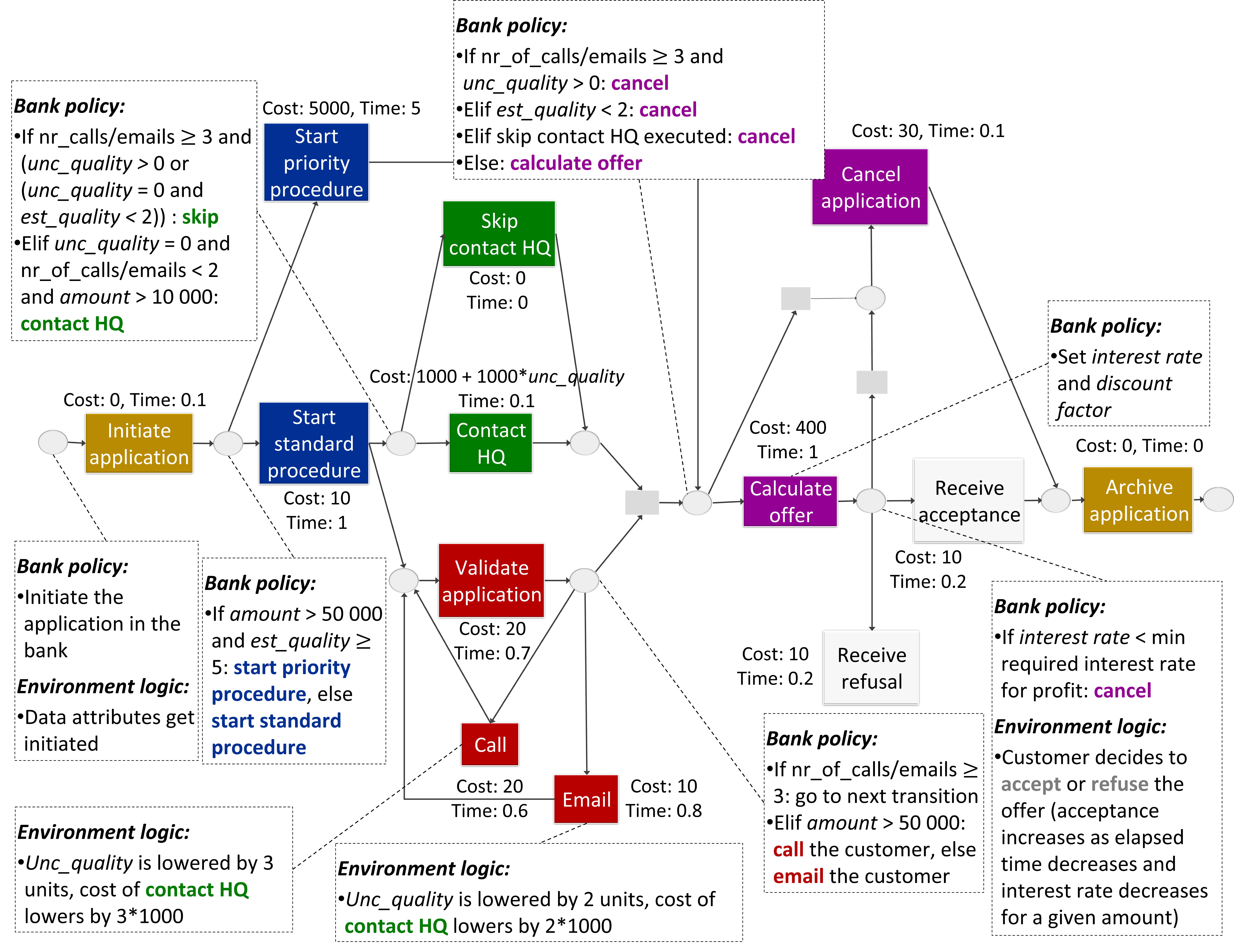}
    \caption{The simulator process model along with the bank policy and environment logic.}
    \label{fig:process}
\end{figure}

The design of the simulator's process model centers on a fictional loan application process in a bank.
The model is illustrated in Figure~\ref{fig:process}, which includes activity costs, durations, the bank's loan processing policy, and external environmental factors. The process begins with an application submission, followed by the bank selecting a procedure, potentially contacting HQ and the customer, calculating an interest rate, or canceling the application. If an interest rate is offered, the client’s response follows.
Table~\ref{tab:attributes} lists the data attributes guiding the bank policy and the client’s (actual but unobservable) quality, an external factor beyond the bank's control. The target variable to optimize is the loan profit, considering variables like the amount requested, interest rate, elapsed time, costs, (actual) client quality, and cancellation reasons. The client’s decision also affects profit, but is in itself influenced by elapsed time, interest rate, and amount. For simplicity, potential client default is excluded.
We have integrated three main interventions into the loan application process: \textit{Choose procedure}, \textit{Set interest rate}, and \textit{Time contact HQ}. The intervention dimensions and details are outlined in Table~\ref{tab:interventions}.

\begin{table}[!ht]
\fontsize{7}{8}\selectfont
\centering
\caption{Data attributes generated by the simulator. The unobserved attribute 'quality' cannot be used to improve the policy.}
\label{tab:attributes}
\begin{tabular}{l l}
\toprule
\textbf{Attribute} & \textbf{Description} \\ 
\midrule
\textit{Activity} & Executed activity \\ 
\textit{Timestamp} & Activity timestamp \\ 
\textit{Case\_nr} & Case number \\ 
\textit{Cost} & Cumulative cost of the case \\ 
\textit{Amount} & Requested loan amount \\ 
\textit{Est\_quality} & Quality of the client, estimated by the bank \\ 
\textit{Unc\_quality} & Uncertainty of the bank regarding the estimated quality \\ 
\textit{Interest rate} & Interest rate offered to the client \\ 
\textit{Discount factor} & Discount factor, taking into account the estimated quality \\
\cellcolor{gray!20}
\textit{Quality} & Actual quality of the client, unobserved by the bank \\
\bottomrule
\end{tabular}%
\end{table}
\begin{table}
\caption{The main interventions included in the simulator.}\label{tab:interventions}
\setlength{\tabcolsep}{5pt}
\fontsize{7}{8}\selectfont
\centering
\renewcommand\tabularxcolumn[1]{m{#1}}
\begin{tabularx}{\textwidth}{c c c X}
    \toprule
    \textbf{Intervention} & \textbf{\parbox{1cm}{Action\\width}} & \textbf{\parbox{1cm}{Action\\depth}} & \multicolumn{1}{c}{\textbf{Description}}\\
    \midrule
    \textit{Choose procedure} & 2 & 1 & Choose between the standard procedure or the priority procedure. The priority option is faster and skips activities but costs more and ignores client quality uncertainties.\\
    \midrule
    \textit{Set interest rate} & 3 & 1 & Set an appropriate interest rate (0.07, 0.08, or 0.09), balancing profitability with the risk of the client declining the loan offer.\\
    \midrule
    \textit{Time contact HQ} & 2 & 4 & Time the contact with HQ, which runs concurrently with customer contact. Contacting HQ is costly and time-consuming, so it should be initiated early to minimize additional delays alongside customer contact. However, the higher the client's quality uncertainty, the greater the cost of contacting HQ. Since uncertainty lowers after each customer contact, contacting HQ early is not always optimal. Additionally, HQ contact should only ever occur when the bank is confident the application will not be canceled, avoiding excess costs. If HQ contact is not made, it will be skipped, leading to automatic cancellation. Interventions after 'validate application' are excluded, reducing the action depth from 8 to 4, as customer variables remain unchanged during validation, rendering the intervention points directly after validation irrelevant.\\
    \bottomrule
\end{tabularx}
\vspace{-15pt}
\end{table}
\subsection{Implementation and Capabilities}
\subsubsection{Tooling.}
The simulator is built in Python, primarily using the \emph{simpy} library \cite{simpy2017}. While the initial code is based on the PNSIM framework \cite{pnsim2021}, it is extensively modified to suit the requirements of our research, incorporating the specific features discussed below. We use random number generators to generate variables. The code and comprehensive guidance on using SimBank can be found at Zenodo \cite{SimBank}.

\subsubsection{Requirements Realization.}
The first requirement aims to ensure a \underline{complete} \underline{evaluation}. In the simulator, the generative functions are predefined, allowing us to precisely assess the outcomes of every possible action at an intervention point. To maintain consistency across methods, we use a fixed seed for the random number generators. This guarantees that the same recommended actions for a specific test case always leads to the same outcome for that action, both in control-flow and data-flow, enabling a fair comparison of PresPM methods.

The second requirement aims to enable \underline{inter-intervention comparisons}. This is achieved by incorporating the three distinct interventions, each varying in action width and depth. Additionally, the ability to combine these interventions facilitates further research into intervention sequences. SimBank allows users to modify interventions, such as adjusting the logic of an intervention function (e.g., setting the interest rate), as well as introduce new ones---either by adding functions that influence specific variables during an activity or by creating entirely new activities.

The third requirement focuses on enabling \underline{intra-intervention method comparisons}. The simulator allows for generating datasets based on existing bank policies (to test offline methods using fixed datasets) or by randomly selecting actions during interventions. The latter approach mimics an RCT, ensuring unconfounded data collection. By adjusting the proportion of RCT data versus (bank) policy-based data in the dataset, we can control the level of confounding, ranging from $0$ to $1$, which we denote as $\delta$. This enables us to assess the robustness of offline methods against confounding introduced by existing policies across a specified range. The bank policy can also be adjusted to examine how the performance of the data-gathering policy impacts method performance. Additionally, we support comparisons between offline and online methods. The simulator operates in two main modes: offline simulation, which returns a complete event log, and online mode. In online mode, also used for method evaluation, a case is generated up to a specified intervention point where an external policy dictates the action to be taken. Afterwards, the case simulation resumes until the next potential intervention point, repeating this cycle. If no further intervention points exist, the simulation continues until the outcome can be calculated. 

The final requirement aims to \underline{balance realism with simplicity}, which we achieve in several implementation choices. SimBank includes true activity concurrency, previously not present in the PNSIM simulator. This advancement enables activities in different parallel branches to be executed before, after, or simultaneously with one another, making the simulator more realistic. We accomplish this by designating which activities belong to the various parallel branches and by explicitly setting the timestamps to indicate whether an activity occurs before, during, or after an activity in another branch. Some activities are also arranged into loops, as shown in Figure \ref{fig:process}. This greatly enhances the structural diversity of the process, particularly when compared to the simpler configurations outlined in \cite{weytjens2023}, which remains the only other study employing a fully synthetic setup. The underlying logic of SimBank draws inspiration from real-world banking practices. For example, the risk-free rate, a key part of the discount rate, is modeled after the 10-year Belgian government bond yield at the time of the simulator's development. Profit, the target variable, is calculated using the net present value, reflecting the time-related value of money~\cite{dahlquist2022}. The activities in SimBank are also based on the most commonly used event logs from PresPM: BPIC12 and BPIC17. Examples of activities that appear in both SimBank and these logs include \textit{call}, \textit{validate application}, \textit{calculate offer}, \textit{initiate application}, and \textit{receive acceptance/refusal}. At the same time, we intentionally maintain simplicity to ensure the datasets remain interpretable. This includes aggregating complex factors, such as client quality, into single variables and limiting the overall number of activities. While we incorporate more activities than in \cite{weytjens2023}, we avoid excessive loops or overly complex parallel structures, maintaining a balance that supports clear conclusions. If users feel the process is still too simplified, they have the option to add more activities in the same way as the existing ones.

\section{Proof of Concept}\label{sec:poc}
This section focuses on validating SimBank for PresPM benchmarking, demonstrating its ability to highlight method differences, support both online and offline approaches, and enable diverse experimental scenarios. we emphasize the importance of SimBank’s accurate evaluation, which stands out compared to existing evaluation methods. The focus is on showcasing SimBank's capabilities rather than optimizing a PresPM method's performance. We begin with the experimental setup and then present the results.

\subsection{Experimental Setup}
We experiment with 4 PresPM methods---2 based on CI and two based on RL---to assess their effectiveness in optimizing interventions. We also use the widely adopted RealCause evaluation method to compare its approximated evaluation results with the accurate SimBank evaluation. First, we explain the RealCause implementation, followed by the 4 PresPM methods.
\begin{itemize}
    \item \textbf{Evaluation method: RealCause.} RealCause was developed to benchmark CI approaches by generating counterfactuals that reflect a real dataset, helping to assess whether a CI method recommends the optimal actions. RealCause has been used in PresPM research, such as in \cite{bozorgi2023CI,bozorgi2023RL,shoush2024,shoush2022,shoushwhitebox}. Given the intervention action recommended by the PresPM approach for a test case, RealCause generates the corresponding outcome. The method models two key distributions. The first is the selection distribution, which determines which intervention action is assigned to a case and is modeled here by a Bernoulli distribution for interventions \textit{Choose procedure} and \textit{Time contact HQ}, and a categorical distribution for \textit{Set interest rate}. The second is the outcome distribution, modeled by a Sigmoid Flow distribution with discrete value concentrations \cite{huang2018neural}. The distributions are approximated using a TARNet-based structure, which includes a neural network to learn treatment-agnostic representations and two sub-networks tailored to specific treatment groups \cite{shalit2017}. This design could be effective as it utilizes all data for the initial representation network while emphasizing treatment through the separate networks. The models were tested using the statistical tests of the original RealCause paper \cite{realcause2020}.
    \item \textbf{PresPM method 1: Standard S-learner (CI).} As a standard CI implementation, we adopt the established S-learner framework from Causal Machine Learning \cite{kunzel2017}, where a single model is trained with the intervention action as an input variable. We employ an LSTM network, which takes as input the case variables, actions, and event variables of a prefix up to an intervention point. The output is the predicted outcome (profit) for the given prefix and current action. For an intervention sequence, we can extend this approach by training separate models for each intervention. Preprocessing includes one-hot encoding for categorical features, standardization of numeric variables, and sequence padding. We use early-stopping during training to prevent overfitting. For interventions with action depth 1 (fixed), the optimal action is chosen by predicting and comparing outcomes across all possible actions. For interventions with an action depth larger than 1 (such as \textit{Time contact HQ}), the intervention is triggered when the predicted profit difference between intervening and not intervening exceeds a threshold determined on a validation set. This approach is referred to as the standard S-learner.
    \item \textbf{PresPM method 2: RealCause-based S-learner (CI).} The second CI approach is inspired by \cite{bozorgi2023RL} and \cite{bozorgi2023CI}, and combines RealCause with the S-learner framework. RealCause is first used to generate counterfactuals by estimating profits for actions other than those observed in the real training dataset (from SimBank). We then train an XGBoost model instead of an LSTM, as RealCause requires non-sequential encoding. Preprocessing follows the approach in \cite{bozorgi2023CI}, which includes aggregation encoding. A threshold for timed interventions is again determined on a validation set. In summary, this approach uses RealCause as a data augmentation tool, as in \cite{bozorgi2023RL}, while integrating it with a CI framework, as in \cite{bozorgi2023CI}. We refer to this CI approach as RealCause-based S-learner.
    \item \textbf{PresPM method 3: Deep Q-learning (RL).} We adopt the DQN approach as standard RL implementation, employing two neural networks---an online network and a target network---for training stability. Both networks share the same LSTM architecture as the standard S-learner for consistency. During inference, the online network processes the current state, including all prior event variables, case variables, and actions, to output Q-values for each possible action. The action with the highest Q-value is chosen. Rewards are based on the case's final profit, with penalties for interventions at non-intervention points. We refer to this RL method as Deep Q-learning.
    \item \textbf{PresPM method 4: K-means-based Q-learning (RL).} Our second RL approach is inspired by \cite{branchi2022}. Their method involves preprocessing prefixes by analyzing activity counts and positions, then using K-means clustering to group them. An event log is replayed to build an RL training environment, where states are defined by the prefix cluster and the last activity. In our implementation, we simplify this by using only K-means clustering and the state representation constructed from an offline dataset. Traditional Q-learning is applied to generate a Q-table online using cases from SimBank, with the same reward structure as the Deep Q-learning approach. Unlike \cite{branchi2022}, we skip replaying an event log to construct the environment. We refer to this approach as K-means-based Q-learning.
\end{itemize}

To provide a general overview of SimBank’s capabilities, we begin by applying the standard PresPM methods (S-learner and DQN) across all interventions. Additionally, we apply the standard S-learner to an intervention sequence combining \textit{Choose procedure} and \textit{Set interest rate}, but exclude this for DQN due to its high computational demands. Next, we extend our analysis for the \textit{Time contact HQ} intervention, as its dimensions are the most studied in PresPM research (action width 2 and action depth $>$ 1, see Section \ref{subsec:background_limitations}). For this intervention, we incorporate the methods inspired by existing approaches: the RealCause-based S-learner and K-means-based Q-learning. Finally, to assess the relevance of the confounding parameter in SimBank, we adjust $\delta$ for offline methods (CI-based and RealCause), testing two values for \textit{Time contact HQ} ($\delta = 0$ and $\delta = 0.999$), and six levels for the standard S-learner. All other experiments use $\delta = 0$.

All methods are trained on 100,000 cases, with 1,000 for validation and 10,000 for testing. RealCause and the K-means model are trained on an extra unrelated dataset of 10,000 cases. The evaluation metric, denoted as \textit{Gain}, is the relative change in total profit compared to the bank's baseline policy. A policy therefore outperforms the bank policy if its $\textit{Gain}> 0$. Each experiment is repeated 5 times. 

\subsection{Results}

Table \ref{tab:other_interventions} presents the results for the \textit{Choose procedure} and \textit{Set interest rate} interventions. Note that the performance of the bank policy is not in the table, as its \textit{Gain} is naturally 0. In both cases, the standard S-learner and Deep Q-learning outperform the random policy and bank policy. For the \textit{Choose procedure} intervention, the standard S-learner demonstrates the best performance and greater stability, with a lower standard deviation. For the \textit{Set interest rate} intervention, the standard S-learner also slightly outperforms Deep Q-learning and remains the more stable method. While Q-learning theoretically offers optimal solutions \cite{sutton2018}, its practical implementation underperforms. Potential improvements include exploring alternative state representations, architectures, or efficient sampling techniques \cite{effsamplingRL2022}.
\begin{table}
    \centering
    \fontsize{8}{9}\selectfont
    \caption{Comparison of approaches for interventions \textit{Choose procedure} and \textit{Set interest rate}, using a training set for CI generated with no confounding ($\delta = 0$).}
    \label{tab:other_interventions}
    \setlength{\tabcolsep}{5pt}
    \begin{tabular}{l l c c c c}
        \toprule
            \multicolumn{1}{l}{Approach} & \multicolumn{1}{l}{Set-up} & \multicolumn{2}{c}{\textit{Choose procedure}} & \multicolumn{2}{c}{\textit{Set interest rate}}\\
                \cmidrule(r){3-4} \cmidrule(l){5-6}
             & & \multicolumn{1}{c}{Gain} & \multicolumn{1}{c}{StDev} & \multicolumn{1}{c}{Gain}  & \multicolumn{1}{c}{StDev}\\
            \midrule
            \multicolumn{1}{l}{Random} & - & -0.209 & 0.008 & -0.297 & 0.009\\
            \multicolumn{1}{l}{CI} & \multicolumn{1}{l}{Standard S-learner} & \textbf{0.445} & 0.011 & \textbf{0.144} & 0.001\\
            \multicolumn{1}{l}{RL} & \multicolumn{1}{l}{Deep Q-Learning} & 0.323 & 0.015 & 0.131 & 0.007\\
        \bottomrule
    \end{tabular}
\end{table}

Table \ref{tab:time_contact_HQ_unbiased} shows results for the \textit{Time contact HQ} intervention on confounding-free datasets ($\delta = 0$), matching an RCT. All methods outperform the random and bank policies, with Deep Q-learning achieving the highest gain. The standard S-learner performs less effectively and is less stable, aligning with the findings in \cite{weytjens2023}. Methods inspired by prior works perform worse than standard implementations. The RealCause-based S-learner likely loses temporal information due to aggregation encoding and accumulates errors due to dual-model use (RealCause and XGBoost). Similarly, K-means-based Q-learning likely sacrifices temporal detail by solely relying on cluster labels and the last activity for state representation. Interestingly, method rankings are consistent between the RealCause and true evaluations, as the RCT training data ensures accurate counterfactuals. However, RealCause underestimates absolute performance.
\begin{table}
    \centering
    \fontsize{8}{9}\selectfont
    \caption{Comparison of approaches for intervention \textit{Time contact HQ}, using a training set for CI and RealCause with no confounding ($\delta = 0$).}
    \label{tab:time_contact_HQ_unbiased}
    \setlength{\tabcolsep}{5pt}
    \begin{tabular}{p{1ex} p{2cm} c c c c}
        \toprule
            \multicolumn{2}{l}{Approach \& Set-up} & \multicolumn{2}{c}{True} & \multicolumn{2}{c}{RealCause}\\
                \cmidrule(r){3-4} \cmidrule(l){5-6}
             & & \multicolumn{1}{c}{Gain} & \multicolumn{1}{c}{StDev} & \multicolumn{1}{c}{Gain}  & \multicolumn{1}{c}{StDev}\\
            \midrule
            \multicolumn{2}{l}{Random} & -0.235 & 0.016 & -0.230 & 0.033\\
            \multicolumn{2}{l}{CI}\\
            & \multicolumn{1}{l}{Standard S-learner} & 0.691 & 0.072 & 0.619 & 0.075\\
            & \multicolumn{1}{l}{RealCause-based S-learner} & 0.565 & 0.001 & 0.457 & 0.101\\
            \multicolumn{2}{l}{RL}\\
             & \multicolumn{1}{l}{Deep Q-Learning} & \textbf{0.766} & 0.013 & \textbf{0.675} & 0.064\\
            & \multicolumn{1}{l}{K-means-based Q-Learning} & 0.661 & 0.037 & 0.584 & 0.055\\
        \bottomrule
    \end{tabular}
\end{table}


Table \ref{tab:time_contact_HQ_biased} presents results for the \textit{Time contact HQ} intervention with $\delta = 0.999$, creating a heavily confounded training dataset. Note that the performance of the random policy remains unchanged from the unconfounded setting, since it does not rely on the training data. All methods outperform the random and bank policies, though the standard S-learner only marginally exceeds the bank policy. True evaluation shows the RealCause-based S-learner performs best, while the standard S-learner is more stable. However, performance suffers across the board, as the dataset is confounded and lacks diversity. The RealCause-based S-learner performs better, likely due to the TARNet-based structure in RealCause which emphasizes treatment selection. Moreover, the RealCause evaluation model leads to inaccurate evaluations when trained on a confounded dataset. It overestimates the RealCause-based S-learner, due to the alignment between this learner and the RealCause evaluation model. This underscores the risk of relying on RealCause for both data augmentation during training and for defining evaluation metrics, as seen similarly in \cite{bozorgi2023RL}, albeit with an RL agent instead. More critically, RealCause ranks the random policy higher than the standard S-learner, despite the true evaluation indicating otherwise. These findings emphasize the importance of accurate and unbiased evaluation methods, such as those provided by SimBank. Since most real-world event logs are collected under existing, non-randomized policies (unlike RCTs), they are likely to be confounded. In such cases, relying solely on RealCause can lead to misleading conclusions.
\begin{table}
    \centering
    \fontsize{8}{9}\selectfont
    \caption{Comparison of approaches for intervention \textit{Time contact HQ}, using a training set for CI and RealCause with strong confounding ($\delta = 0.999$).}
    \label{tab:time_contact_HQ_biased}
    \setlength{\tabcolsep}{5pt}
    \begin{tabular}{p{1ex} p{2cm} c c c c}
        \toprule
            \multicolumn{2}{l}{Approach \& Set-up} & \multicolumn{2}{c}{True} & \multicolumn{2}{c}{RealCause}\\
                \cmidrule(r){3-4} \cmidrule(l){5-6}
             & & \multicolumn{1}{c}{Gain} & \multicolumn{1}{c}{StDev} & \multicolumn{1}{c}{Gain}  & \multicolumn{1}{c}{StDev}\\
            \midrule
            \multicolumn{2}{l}{Random} & -0.235 & 0.016 & 0.091 & 0.640\\
            \multicolumn{2}{l}{CI}\\
            & \multicolumn{1}{l}{Standard S-learner} & 0.053 & 0.375 & -0.008 & 0.858\\
            & \multicolumn{1}{l}{RealCause-based S-learner} & \textbf{0.198} & 0.454 & \textbf{2.249} & 4.551\\
        \bottomrule
    \end{tabular}
\end{table}

\begin{figure}[h]
\vspace{-15pt}
    \centering
    \includegraphics[width=0.7\textwidth, height=0.45\textwidth]{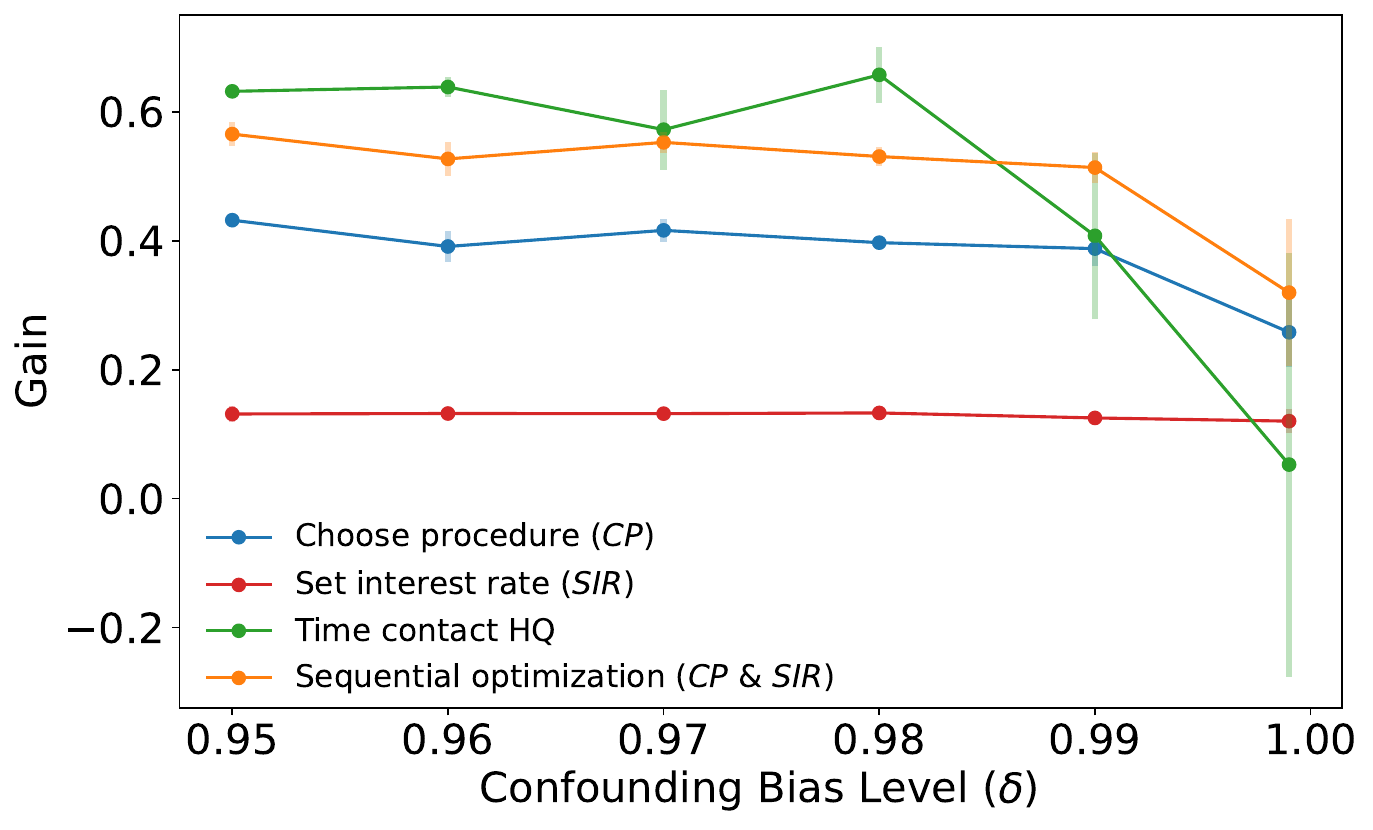}
    \caption{Performance of the standard S-learner across 6 levels of $\delta$ for the 3 main interventions and 1 intervention sequence. A clear trend emerges, showing decreasing performance as $\delta$ increases.}
    \label{fig:varying_delta_no_sum}
\end{figure}

Figure \ref{fig:varying_delta_no_sum} contains the results of the experiments on the standard S-learner where we vary $\delta$ at 6 different values. The minimum value of $0.95$ is chosen since performance stays more or less consistent between $0$ and $0.95$ for a dataset of 100,000 cases. The error bars indicate a confidence interval of the performance at each level. Except for the \textit{Set interest rate} intervention, a general trend is that both performance and stability decrease for higher values of $\delta$. Especially the bank’s original policy for the \textit{Time contact HQ} intervention seems undermining, given the high sensitivity of the standard S-learner to $\delta$. For the sequence of \textit{Choose procedure} and \textit{Set interest rate}, sequential optimization proves beneficial, as our setup outperforms optimizing only one of the interventions. However, since our implementation consists of one model per intervention, each model optimizes its intervention independently, assuming the bank policy for the other. This likely limits overall performance. More research is needed to improve the optimization of intervention sequences, which is especially relevant in business processes.

These results show that using SimBank for method comparisons offers valuable insights. By leveraging SimBank's synthetic data, we can pinpoint the strengths and weaknesses of both online and offline methods. This analysis can span various intervention dimensions and scenarios where challenges might emerge from data-gathering policies. The findings also highlight the critical need for accurate evaluation in PresPM. We have shown that evaluation methods like RealCause can be validated using SimBank.

\section{Conclusion and Future Work}\label{sec:conclusion}
We introduce SimBank, a synthetic data simulator designed to benchmark methods in PresPM research. Current research in this field often lacks comprehensive comparisons and accurate evaluations, and SimBank aims to address these gaps. SimBank simulates a bank's loan application process and supports both online and offline method comparisons. It includes three key intervention problems relevant to business processes, with possible combinations of interventions, and allows researchers to adjust data-generating parameters like confounding levels to create experiments tailored to PresPM research. In our proof of concept, we demonstrate that SimBank can provide valuable insights by enabling comparisons between interventions of varying complexity and between different setups for the same intervention. We also emphasize the importance of precise evaluation, which SimBank facilitates by allowing for fully accurate assessments.

A limitation of our simulator is that SimBank is only partially realistic. While a fully synthetic simulator can never completely replicate real-world scenarios, we have tried to minimize this limitation by incorporating features like true concurrency, loops, and basing it on real-life logs and variables. Another limitation is the assumption that there are no interdependencies between process instances. However, this can be addressed in future work, as we used PNSIM as our starting point, which can potentially account for, e.g., delays in one process instance caused by others.

Future research will focus on optimizing intervention sequences, particularly for offline methods like CI. The approach outlined in \cite{bica2020} shows promise in this regard, but requires further investigation to determine its effectiveness in highly variable and complex business process environments. Additionally, we plan to extend the simulator with case interdependencies by including the potential delays and resource limitations. Lastly, Causal Discovery (CD) for PresPM can be explored to identify interventions~\cite{causaldiscovery2023}. CD methods can be tested with SimBank. Interventions and their dimensions are often assumed predefined in PresPM, which may not always apply in real-world business processes.

\subsubsection{Acknowledgments}
This work was supported the Research Foundation Flanders (FWO) under grant number G039923N and 11A6J25N, and Internal Funds KU Leuven under grant number C14/23/031.

\bibliographystyle{splncs04}
\bibliography{references}

\end{document}